\pdfoutput=1
\RequirePackage{ifpdf}
\ifpdf 
\documentclass[pdftex]{sigma}
\else
\documentclass{sigma}
\fi

\begin{document}

\allowdisplaybreaks

\renewcommand{\PaperNumber}{083}

\FirstPageHeading

\ShortArticleName{Three-Qubit Observables and Hyperplanes of the Smallest Split Cayley Hexagon}

\ArticleName{`Magic' Conf\/igurations of Three-Qubit Observables\\ and Geometric Hyperplanes of the Smallest Split\\ Cayley Hexagon}

\Author{Metod SANIGA~$^{\dag^1}$, Michel PLANAT~$^{\dag^2}$, Petr PRACNA~$^{\dag^3}$ and P\'eter L\'EVAY~$^{\dag^4}$}

\AuthorNameForHeading{M.~Saniga, M.~Planat, P.~Pracna and P.~L\'evay}

\Address{$^{\dag^1}$~Astronomical Institute, Slovak Academy of Sciences,\\
\hphantom{$^{\dag^1}$}~SK-05960 Tatransk\' a Lomnica, Slovak Republic}
\EmailDD{\href{mailto:msaniga@astro.sk}{msaniga@astro.sk}}

\Address{$^{\dag^2}$~Institut FEMTO-ST, CNRS,
32 Avenue de
l'Observatoire, F-25044 Besan\c con Cedex, France}
\EmailDD{\href{mailto:michel.planat@femto-st.fr}{michel.planat@femto-st.fr}}

\Address{$^{\dag^3}$~J. Heyrovsk\' y Institute of Physical
Chemistry, v.v.i.,  Academy of Sciences\\
\hphantom{$^{\dag^3}$}~of the Czech Republic, Dolej\v{s}kova 3, CZ-182 23 Prague 8, Czech
Republic}
\EmailDD{\href{mailto:pracna@jh-inst.cas.cz}{pracna@jh-inst.cas.cz}}

\Address{$^{\dag^4}$~Department of Theoretical Physics, Institute of Physics,\\
\hphantom{$^{\dag^4}$}~Budapest University of Technology and Economics,
H-1521 Budapest, Hungary}
\EmailDD{\href{mailto:levay@neumann.phy.bme.hu}{levay@neumann.phy.bme.hu}}

\ArticleDates{Received June 22, 2012, in f\/inal form November 02, 2012; Published online November 06, 2012}

\Abstract{Recently Waegell and Aravind [\textit{J.~Phys.~A: Math. Theor.} \textbf{45} (2012),
  405301, 13~pages] have given a number of distinct sets of three-qubit observables, each furni\-shing a proof of the Kochen--Specker theorem. Here it is demonstrated that two of these sets/conf\/igurations, namely the $18_{2} - 12_{3}$ and $2_{4}14_{2} - 4_{3}6_{4}$ ones,  can uniquely be extended into geometric hyperplanes of the split Cayley hexagon of order two, namely into those of types ${\cal V}_{22}(37; 0, 12, 15, 10)$ and ${\cal V}_{4}(49; 0, 0, 21, 28)$ in the classif\/ication of Frohardt and Johnson [\textit{Comm. Algebra} \textbf{22} (1994), 773--797]. Moreover, employing an automorphism of order seven of the hexagon, six more replicas of either of the two conf\/igurations are obtained.}

\Keywords{`magic' conf\/igurations of observables; three-qubit Pauli group; split Cayley hexagon of order two}

\Classification{51Exx; 81R99}

\section{Introduction}
For a relatively long time, the only known `magic' conf\/iguration of three-qubit observables, that is a conf\/iguration furnishing a proof of the Kochen--Specker theorem \cite{ks}, was the so-called Mermin pentagram~\cite{mer}~-- an aggregate of ten observables forming f\/ive sets of four mutually commuting elements each,  such that each observable belongs to two of these sets and the product of observables in one of them is the minus identity, whilst in the remaining four it is the plus identity. Very recently,  Waegell and Aravind~\cite{wa} have proposed a procedure that generates a~variety of such conf\/igurations.  The purpose of this note is to provide the reader, following the spirit and strategy of two recent papers~\cite{sl,sp12},  with an intriguing f\/inite-geometric insight into a couple of them. The relevant f\/inite geometry is here that of the split Cayley hexagon of order two~\cite{fj,psm,schr}, this being a distinguished subgeometry of the symplectic  polar space $W(5,2)$
associated with the three-qubit Pauli group \cite{hos,pla, ps,sp07,spp,thas}, and two (out of as many as~25) distinct types of its geometric hyperplanes.

To be more explicit, we shall make use of a highly symmetric rendering/f\/igure \cite{psm, schr} of the split Cayley hexagon of order two, where each of its 63 points  is associated with one of 63 non-trivial elements of the three-qubit Pauli group in such a way that the product of any three elements represented by the points on the same line will be proportional to the identity~\cite{lsv}. Then, by embedding in the hexagon (which amounts to highlighting in the f\/igure) each of the two above-mentioned Waegell--Aravind magic conf\/igurations, we shall diagrammatically illustrate consecutive steps of `line-completion' that in either case lead to a unique geometric hyperplane of the particular type. In addition, an automorphism of order seven of the f\/igure in question will, in either case as well, be shown to give birth to six more conf\/igurations having the same `magic' nature as the original one.

This short contribution is organized as follows. Section~\ref{section2} highlights rudiments of the three-qubit Pauli group and its associated symplectic polar space $W(5,2)$, introduces the split Cayley hexagon of order two and lists basic properties of all 25 types of its geometric hyperplanes. Section~\ref{section3} deals with two particular three-qubit magic conf\/igurations of Waegell and Aravind~\cite{wa} and presents a detailed demonstration of their completion into specif\/ic geometric hyperplanes of the hexagon. Finally, Section~\ref{section4} summarizes main f\/indings, mentions a parallel with the two-qubit Mermin(--Peres) magic square and outlines some prospective work.

\section{Three-qubit Pauli group and split Cayley hexagon\\ of order two}\label{section2}

The (generalized) three-qubit Pauli group, ${\cal P}_3$, is generated by three-fold tensor products of the matrices
\begin{gather*}
I = \begin{pmatrix}
1 & 0 \\
0 & 1
\end{pmatrix},\qquad
X = \begin{pmatrix}
0 & 1 \\
1 & 0
\end{pmatrix}, \qquad
Y = \begin{pmatrix}
0 & -i \\
i & 0
\end{pmatrix}
\qquad {\rm and} \qquad
Z = \begin{pmatrix}
1 & 0 \\
0 & -1
\end{pmatrix}.
\end{gather*}
Explicitly,
\begin{gather*}
{\cal P}_3 = \big\{i^{\alpha} A_1 \otimes A_2 \otimes A_3:~ A_j \in \{I, X, Y, Z \},~ j \in \{1, 2, 3\},~\alpha \in \{0, 1, 2, 3\} \big\}.
\end{gather*}
Here, we will be dealing with its factored version $\overline{{\cal P}}_3 \equiv {\cal P}_3/{\cal Z}({\cal P}_3)$, where the center ${\cal Z}({\cal P}_3)$ consists of $\pm I \otimes I \otimes I$ and $\pm i I \otimes I \otimes I$,\footnote{In what follows, we shall use a shorthand notation for the tensor product: $A_1 \otimes A_2 \otimes A_3 \equiv A_1 A_2 A_3$.} and whose geometry is that of the symplectic polar space $W(5,2)$ \cite{hos,pla, ps,sp07,spp,thas}. This space, freely speaking, is a collection of all totally isotropic subspaces of the ambient f\/ive-dimensional binary projective space, PG$(5,2)$, equipped with a non-degenerate alternating bilinear form.
The 63 elements of the group are in a bijective correspondence with the 63 points of $W(5,2)$ in such a way that two commuting elements correspond to two points joined by a totally isotropic line; a maximum set of mutually commuting elements of the group having its counterpart in a maximal totally isotropic subspace (also called a generator), which is a projective plane of order two, the Fano plane.
We shall, however, not be concerned with the full geometric structure of $W(5,2)$, but~-- as already mentioned~-- restrict ourselves to its important subgeometry represented by the split Cayley hexagon of order two. Although the two structures are identical as point-sets, the hexagon contains only 63 lines, which is much less than $W(5,2)$, and it is thus more handy to work with.

\begin{figure}[t]\centering
\centerline{\includegraphics[width=11cm]{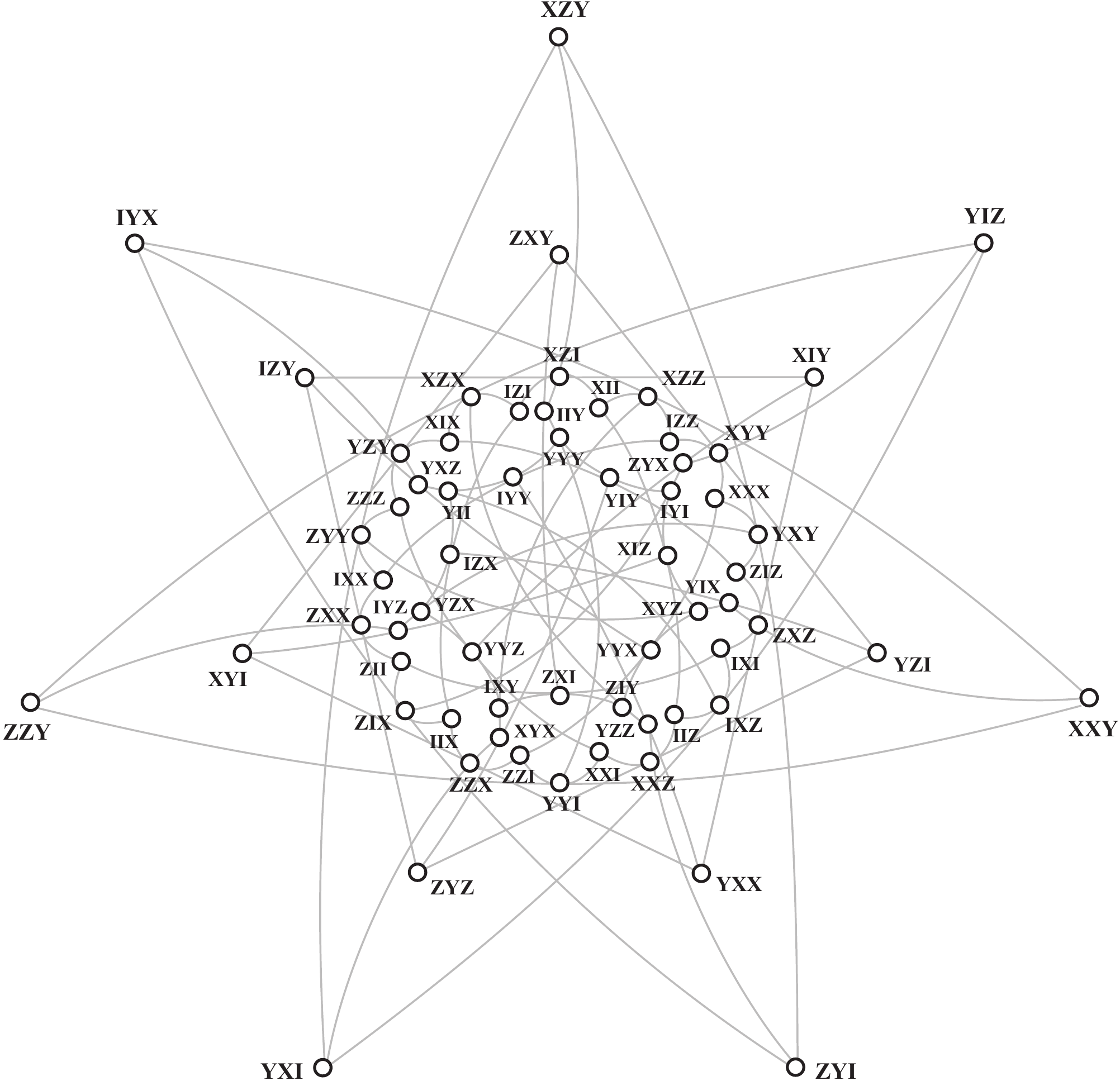}}

\caption{A diagrammatic illustration of the structure of the split Cayley hexagon of order two (based on drawings given in~\cite{psm, schr}). The points are illustrated by small circles and its lines by segments of straight-lines and/or arcs; note that there are many intersections of segments that do not represent any points of the hexagon. Labeling by the elements of $\overline{{\cal P}}_3$ is adopted from~\cite{lsv}. Also obvious is an automorphism of order seven of the structure.}\label{Fig1}
\end{figure}

A split Cayley hexagon of order two, ${\cal G}_2$, is a point-line incidence geometry that satisf\/ies the following
axioms \cite{psm, schr}: a)~every line contains three points and every point is contained in three lines; b)~${\cal G}_2$ does not contain any ordinary $k$-gons for $2\leq k < 6$;
c)~given two points, two lines, or a point and a line, there is at least one ordinary hexagon in~${\cal G}_2$ that contains both objects; and d)~it contains the incidence graph
of the Fano plane~\cite{km}.  As ${\cal G}_2$ is rather small, it can easily be
represented in a diagrammatical form, Fig.~\ref{Fig1}, from which all essential features of its geometrical structure can readily be ascertained.

The f\/inal notion that remains to be introduced is that of a
geometric hyperplane. Given any point-line incidence structure, its geometric hyperplane is a subset of the point-set such that every line of the structure either lies fully in the subset,
or shares with it just one point~\cite{ron}; a~point of a geometric hyperplane is called deep if all the lines passing through it are fully contained in the hyperplane. It has been found~\cite{fj} that ${\cal G}_2$ features $2^{14} - 1 = 16\,383$ geometric hyperplanes that fall into 25 distinct types (according to the orbits of its automorphism group) and 13 classes (in terms of the sizes of their point-sets). This classif\/ication is given in Table~\ref{table1}, where we also adopt the Frohardt--Johnson `f\/ive-tuple' notation \cite{fj}, ${\cal V}_{k}(n; n_0, n_1, n_2, n_3)$, meaning that a hyperplane of the $k$-th type, $1 \leq k \leq 25$, is endowed with $n$ points of which $n_s$, $s \in \{0, 1, 2, 3\}$, belong to exactly $s$ lines contained in the hyperplane; thus, $n_3$ is the number of deep points of a hyperplane. It is of some interest to note in passing that there are two distinct types of hyperplanes that have the same f\/ive-tuple, namely ${\cal V}_{24}$ and ${\cal V}_{25}$.

\begin{table}[t]
\centering
\caption{Classes and types of geometric hyperplanes of the split Cayley
hexagon of order two. For each type one gives the size of its point- (`Pts') and line- (`Lns') sets, number of deep points (`DPts'), total number of distinct copies (`Cps') and the stabilizer group (`StGr') of its orbit; for more group-theoretical details, see~\cite{fj}.}\label{table1}

\vspace*{2mm}
\begin{tabular}{||l|l|cccrl||}
\hline \hline
Class & FJ Type & Pts & Lns & DPts & Cps & StGr \\
\hline\hline
\vspace*{-.30cm}
&&&&&&\\
I & $\cal V$$_{2}$(21;21,0,0,0) & 21 & 0 & 0 & 36 & $PGL(2,7)$  \\
\hline
II& $\cal V$$_{7}$(23;16,6,0,1) &23& 3 & 1 & 126 & $(4 \times 4):S_3$   \\
\hline
III& $\cal V$$_{11}$(25;10,12,3,0) &25&6&0&504& $S_4$ \\
\hline
IV & $\cal V$$_{1}$(27;0,27,0,0) & 27 & 9 & 0 & 28 & $X_{27}^{+}:QD_{16}$ \\
& $\cal V$$_{8}$(27;8,15,0,4) & 27 &  9 & 3+1 & 252 & $2 \times S_4$ \\
& $\cal V$$_{13}$(27;8,11,8,0)  & 27 & 8+1 & 0 & 756 & $D_{16}$ \\
& $\cal V$$_{17}$(27;6,15,6,0) & 27 & 6+3&  0  & 1008 & $D_{12}$ \\
\hline
V & $\cal V$$_{12}$(29;7,12,6,4) & 29 & 12 &  4  & 504 & $S_4$   \\
& $\cal V$$_{18}$(29;5,12,12,0) & 29 & 12 & 0 & 1008 & $D_{12}$ \\
& $\cal V$$_{19}$(29;6,12,9,2) & 29 & 12 &  2nc  & 1008 & $D_{12}$   \\
&  $\cal V$$_{23}$(29;4,16,7,2) & 29 & 12 &  2c  & 1512 & $D_{8}$ \\
\hline
VI & $\cal V$$_{6}$(31;0,24,0,7) & 31 & 15 & 6+1 & 63 & $(4 \times 4):D_{12}$ \\
& $\cal V$$_{24}$(31;4,12,12,3) & 31 & 15 & 2+1 & 1512 & $D_8$ \\
& $\cal V$$_{25}$(31;4,12,12,3) & 31 & 15 & 3 & 2016 & $S_3$ \\
\hline
VII & $\cal V$$_{14}$(33;4,8,17,4) & 33 & 18 & 2+2 & 756 & $D_{16}$ \\
& $\cal V$$_{20}$(33;2,12,15,4) & 33 & 18 & 3+1 & 1008 & $D_{12}$  \\
\hline
VIII& $\cal V$$_{3}$(35;0,21,0,14) & 35 & 21 & 14 & 36 & $PGL(2,7)$ \\
& $\cal V$$_{16}$(35;0,13,16,6) & 35 & 21 & 4+2 & 756 & $D_{16}$ \\
& $\cal V$$_{21}$(35;2,9,18,6) & 35 & 21 &  6  & 1008 & $D_{12}$ \\
\hline
IX & $\cal V$$_{15}$(37;1,8,20,8) & 37 & 24 & 8 & 756 & $D_{16}$ \\
& $\cal V$$_{22}$(37;0,12,15,10) & 37 & 24 &6+3+1& 1008 & $D_{12}$ \\
\hline
X & $\cal V$$_{10}$(39;0,10,16,13) & 39 & 27 & 8+4+1 & 378 & $8:2:2$ \\
\hline
XI& $\cal V$$_{9}$(43;0,3,24,16)  &43&33&12+3+1&252&$2 \times S_4$ \\
\hline
XII& $\cal V$$_{5}$(45;0,0,27,18)  &45&36&18&56&$X_{27}^{+}:D_{8}$  \\
\hline
XIII& $\cal V$$_{4}$(49;0,0,21,28) & 49 & 42 & 28 & 36 & $PGL(2,7)$ \\
\hline \hline
\end{tabular}
\end{table}

At this point we have introduced all the necessary f\/inite-geometrical technicalities to be employed
in the next section to analyse some `magic' conf\/igurations of three-qubit observables.

\section[Waegell-Aravind configurations uniquely extendible into geometric hyperplanes]{Waegell--Aravind conf\/igurations uniquely extendible\\ into geometric hyperplanes}\label{section3}

We shall f\/irst deal with the conf\/iguration shown in Fig.~5 of \cite{wa}, bearing there symbol $18_2 - 12_3$. It consists of 18 observables forming 12 sets of mutually commuting elements of size three each, namely:
\begin{alignat*}{4}
&\{IZI, ZZI, ZII\}, &&\{ZII, ZIZ, IIZ\}, &&\{IIZ, IZZ, IZI\},& \\
&\{IXI, XXI, XII\}, &&\{XII, XIX, IIX\}, &&\{IIX, IXX, IXI\}, &\\
&\{IYI, YYI, YII\}, &&\{YII, YIY, IIY\}, &&\{IIY, IYY, IYI\}, &\\
&\{ZZI, XXI, YYI\},\qquad &&\{YIY, XIX, ZIZ\},\qquad &&\{IZZ, IXX, IYY\},&
\end{alignat*}
each of which represents a line in $W(5,2)$.
It is readily verif\/ied that the product of observables in each set is $+III$ except for the last three sets where it is~$-III$. Given this fact and the fact that each observable belongs to exactly two sets, it is impossible to assign the eigenvalue~$+1$ or~$-1$ to each observable in such a way that these obey the same product rules  as the observables themselves~-- this rendering a proof of the Kochen--Specker theorem~\cite{wa}.

\begin{figure}[t]\centering

\includegraphics[width=75mm]{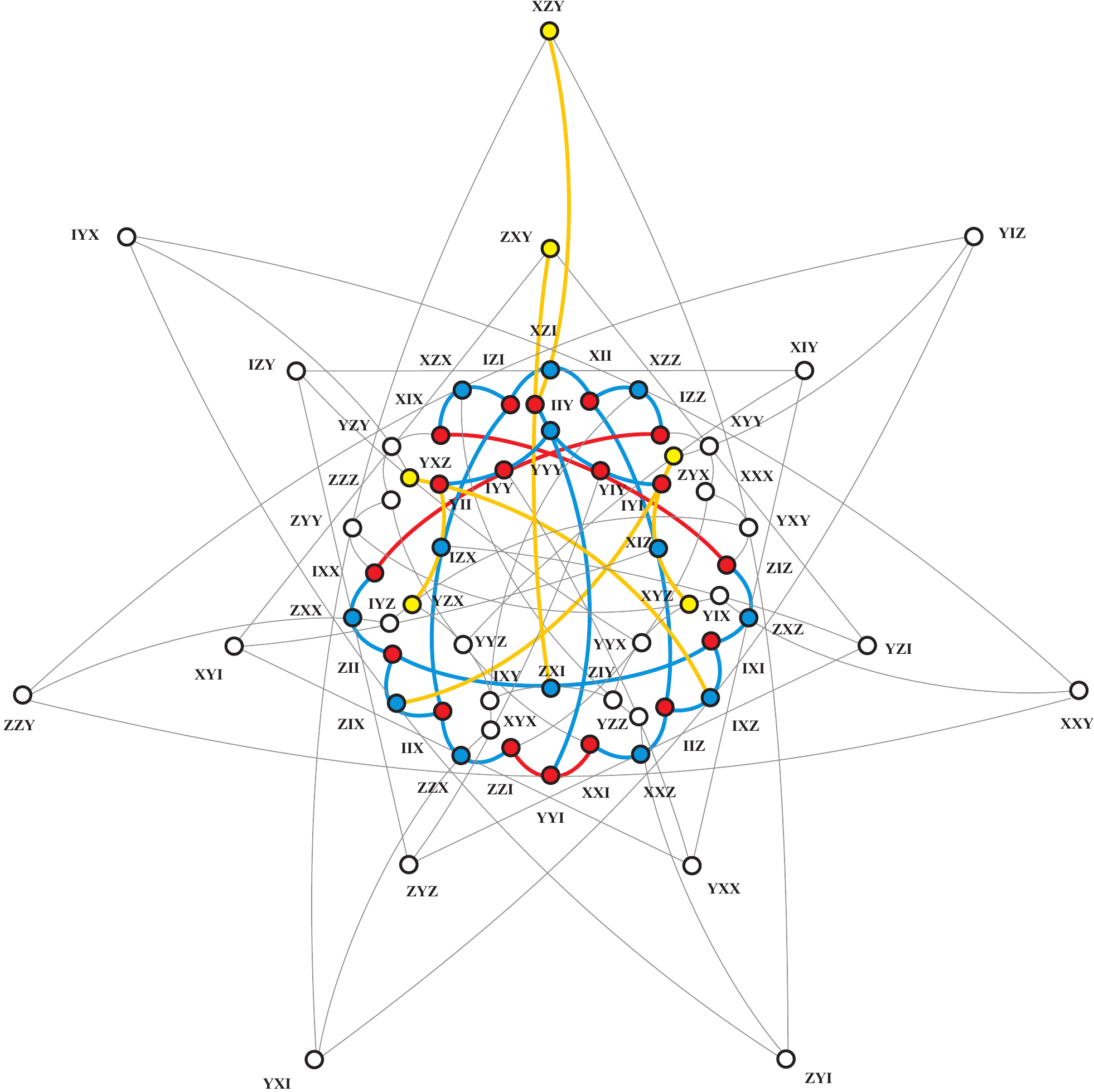}
\qquad
\includegraphics[width=75mm]{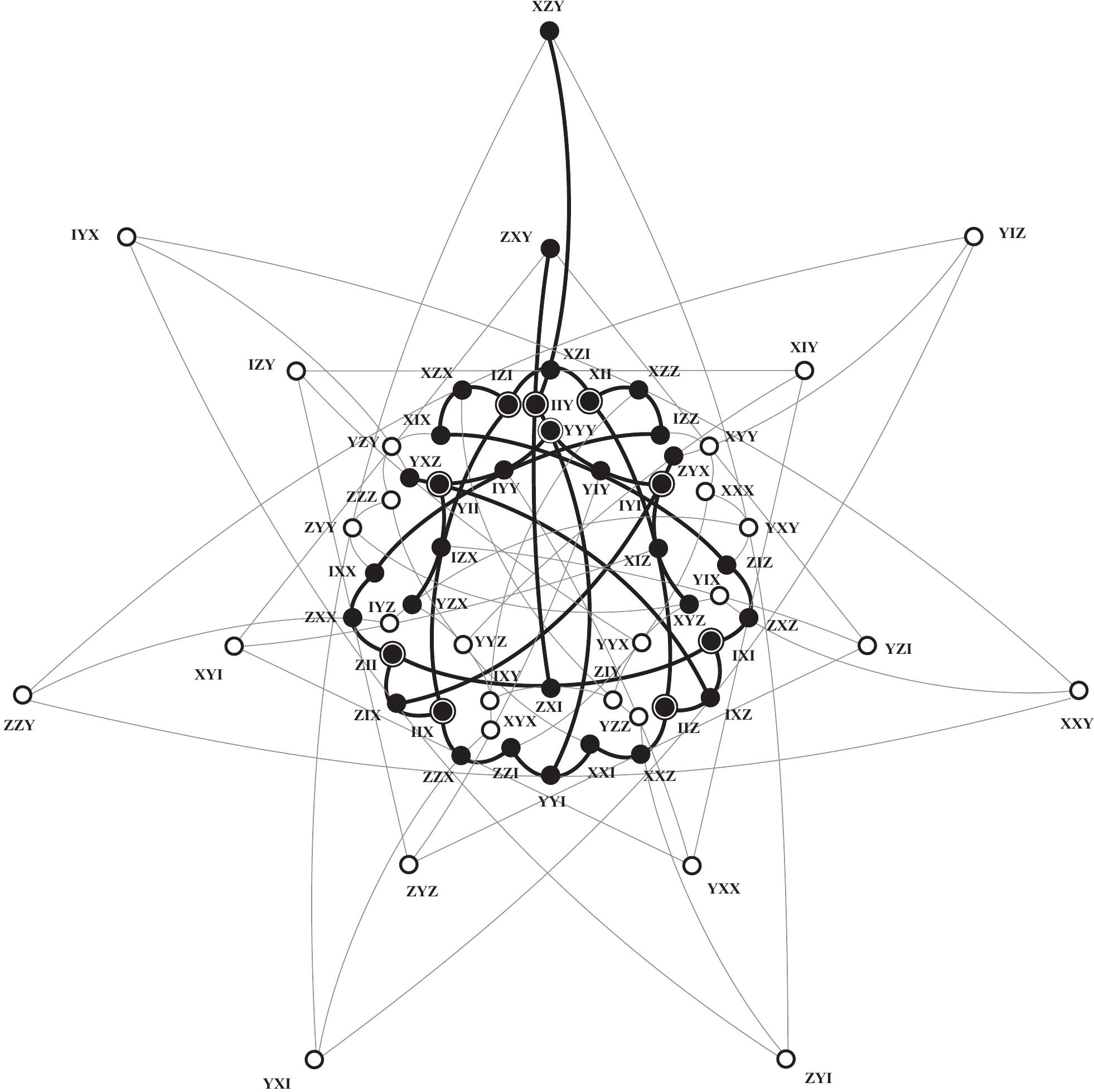}

\caption{{\it Left}:  An illustration of the process of extension of the set of three-qubit observables of the $18_{2} - 12_{3}$ magic conf\/iguration of Waegell and Aravind \cite[Fig.~5]{wa} into a geometric hyperplane of the split Cayley hexagon of order two. Red bullets are the points/observables of the conf\/iguration itself, blue bullets represent the remaining points on the lines of the hexagon featuring two red points and yellow bullets stand for the remaining points on the lines of the hexagon having one red and one blue point. {\it Right}:  A simplif\/ied illustration of the structure of the corresponding hyperplane; encircled are all the ten deep points of the hyperplane.}\label{Fig2}
\end{figure}

It is, however, not this theorem that is of concern for us here. Rather, it is the conf\/iguration {\it as a whole} and, in particular, its image within our split Cayley hexagon, as depicted in Fig.~\ref{Fig2}, {\it left}, by red bullets.
Let us try to f\/ind a geometric hyperplane this conf\/iguration sits in. To this end, one has simply to recall that a line of the hexagon is either {\it fully} contained in a hyperplane, or shares with it a {\it single} point. A brief look at Fig.~\ref{Fig2}, {\it left}, reveals that there are a number of lines of the hexagon that contains two red points and, hence, must lie completely within our sought-for hyperplane; any such line, as well as the remaining third point on it, is illustrated in blue. By extending our original conf\/iguration by the blue points and lines, we shall f\/ind some lines to feature one red and one blue point; these are highlighted in yellow and must equally belong fully to the sought-for hyperplane.  At this step our process of extension ends, for this doubly-extended aggregate of points, illustrated in a colorless version in Fig.~\ref{Fig2}, {\it right}, already satisf\/ies the properties of a geometric hyperplane. From this f\/igure we easily discern that our hyperplane has~37 points and 10 deep points, and from Table~\ref{table1} we f\/ind out that it must be of type ${\cal V}_{22}(37; 0, 12, 15, 10)$, since the only other type of the same size has, for example, only 8 deep points.

\begin{figure}[t]\centering

\includegraphics[width=75mm]{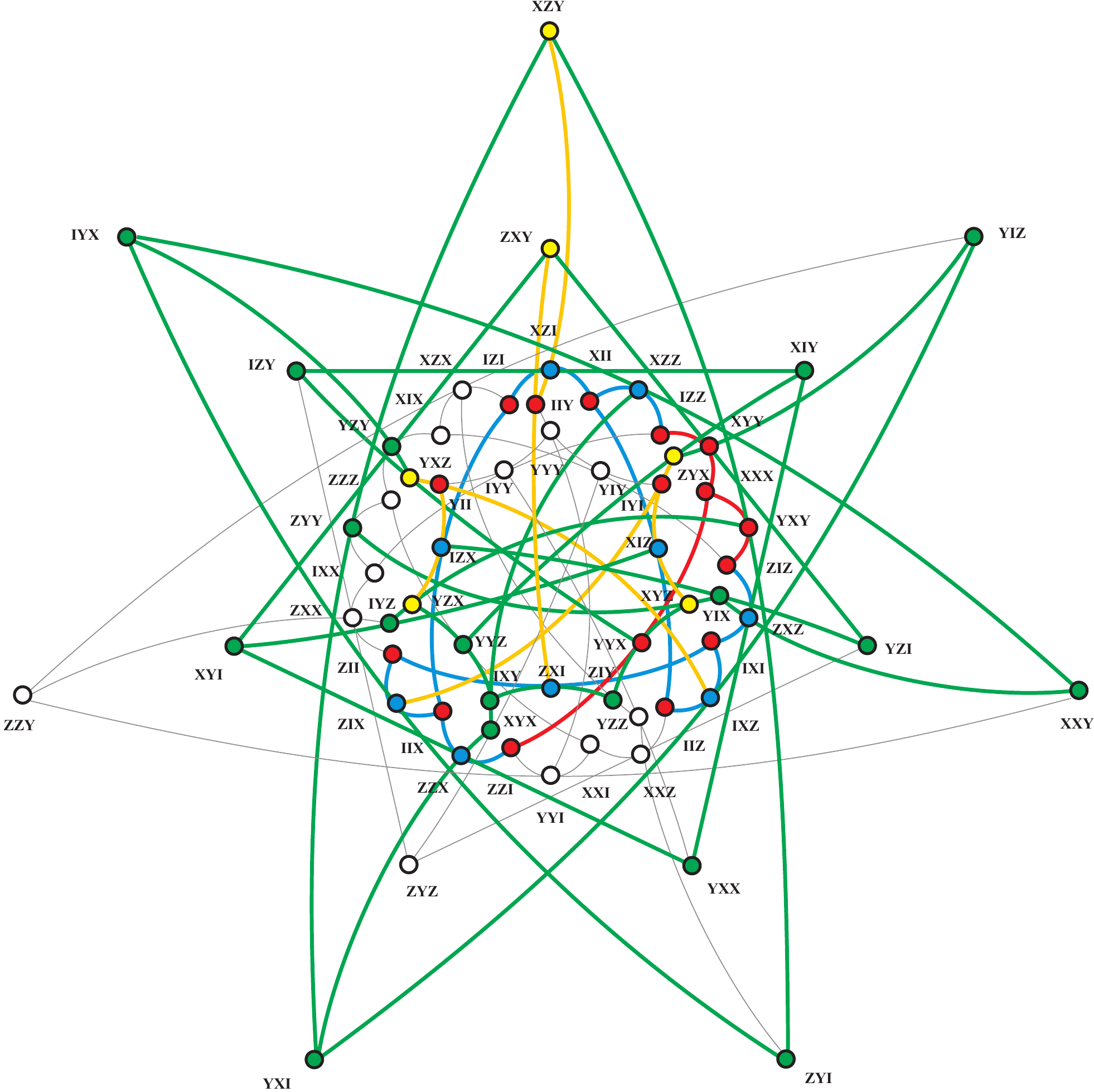}
\qquad
\includegraphics[width=75mm]{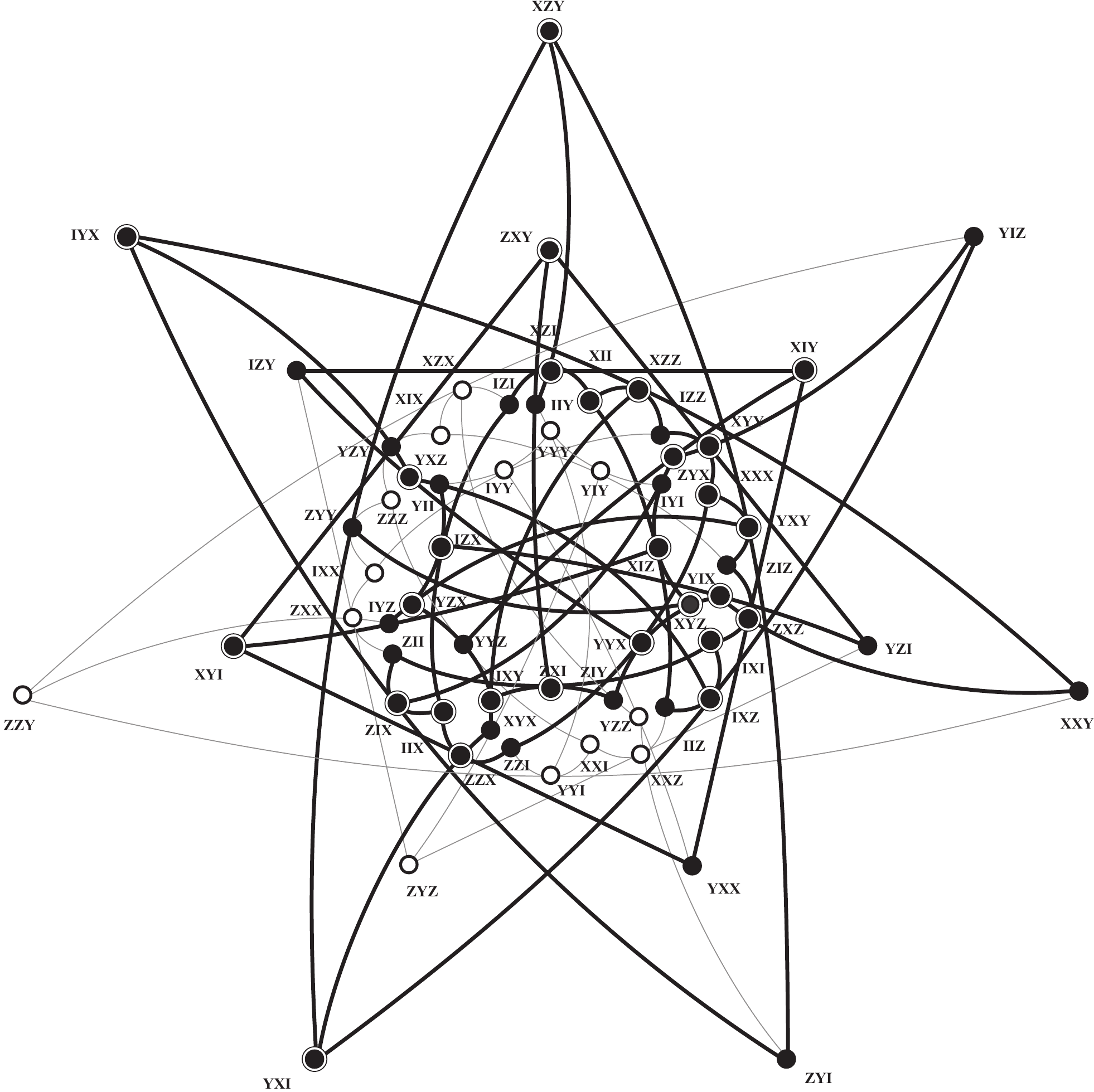}

\caption{{\it Left}:  The same as in Fig.~\ref{Fig2} for the $2_{4}14_{2} - 4_{3}6_{4}$ `magic' conf\/iguration. The meaning of colors is also the same as in the preceding f\/igure save for the fact that we now need a few more steps (for simplicity all illustrated by green color) to arrive at a hyperplane; in particular, a green point is the third point on the line of the hexagon that features either a red point and a yellow one, or a blue/yellow point and a(n already supplied in a previous step) green one.   {\it Right}:
 A simplif\/ied sketch of the resulting hyperplane with its deep points encircled.}\label{Fig3}
\end{figure}

The (only) other W-A conf\/iguration that is subject to a unique extension into a geometric hyperplane is the $2_{4}14_{2} - 4_{3}6_{4}$ one, \cite[Fig.~8, right]{wa}, comprising the following sets of pairwise commuting observables:
 \begin{alignat*}{3}
&\{ZIZ, ZII, IIZ\}, &&\{IIZ, IZI, IZZ\}, &\\
&\{ZII, IZI, ZZI\}, &&\{IZZ, XYY, XXX\}, &\\
&\{IIX, YII, IYI, YYX \}, && \{IYI, IIY, XII, XYY\}, &\\
&\{YII, IIY, IXI, YXY \}, && \{IIX, XII, IXI, XXX\},&\\
&\{IZZ, ZZI, YXY, XXX \}, \qquad && \{IZZ, ZIZ, YYX, XXX\}.&
\end{alignat*}
Geometrically speaking, each three-element set corresponds to a line of $W(5, 2)$, whereas each four-element set represents an af\/f\/ine plane of order two (located in a certain Fano plane)
of $W(5, 2)$. Again, we are interested in the conf\/iguration as a whole and highlighting this set of observables in the f\/igure of the hexagon and following the same strategy as in the preceding case we shall f\/ind that this conf\/iguration is uniquely extendible into
a geometric hyperplane of type ${\cal V}_{4}(49; 0, 0, 21, 28)$~-- see Fig.~\ref{Fig3}. A hyperplane of this type, apart from being of the largest possible size (see Table~\ref{table1}), is also remarkable by the fact that its complement is nothing but the incidence graph of the Fano plane.

As for the remaining W-A conf\/igurations, none of them was found to be uniquely extendible into a geometric hyperplane. Except for the $1_{4}11_{2} - 2_{3}5_{4}$ conf\/iguration \cite[Fig.~8, left]{wa}, which does not seem to be embeddable into any hyperplane, all the other conf\/igurations were found to sit in at least two distinct types of hyperplanes; in particular, the three-qubit Peres--Mermin square~\cite[Fig.~4]{wa} was found in as many as 12 dif\/ferent types of hyperplanes.

We shall conclude this section by observing that an automorphism of order seven of our hexagon can be used to generate six more replicas (listed below in columns  $b$ to $g$) for either of the above
discussed W-A conf\/igurations (column $a$). Technically, this action is performed  by rotating consecutively Fig.~\ref{Fig2}, {\it left}, and Fig.~\ref{Fig3}, {\it left}, by 360/7 degrees around the center of the hexagon.
Thus, rotating the f\/igures in question counterclockwise, the interested reader can easily check that for the $18_{2} - 12_{3}$ case we get
\begin{alignat*}{14}
&~~a \ && \  \mapsto \ && \  ~~b \ && \  \mapsto \ && \  ~~c \ && \  \mapsto \ && \  ~~d \ && \  \mapsto \ && \  ~~e \ && \  \mapsto \ && \  ~~f \ && \  \mapsto \ && \  ~~g, &\\
&XIX \ && \  \mapsto \ && \  IXX \ && \  \mapsto \ && \  IIX \ && \  \mapsto \ && \  XXI \ && \  \mapsto \ && \  IXI \ && \  \mapsto \ && \  XXX \ && \  \mapsto \ && \  XII, &\\
&IZI \ && \  \mapsto \ && \  ZZZ \ && \  \mapsto \ && \  ZII \ && \  \mapsto \ && \  ZZI \ && \  \mapsto \ && \  IIZ \ && \  \mapsto \ && \  ZIZ \ && \  \mapsto \ && \  IZZ, &\\
&XII \ && \  \mapsto \ && \  XIX \ && \  \mapsto \ && \  IXX \ && \  \mapsto \ && \  IIX \ && \  \mapsto \ && \  XXI \ && \  \mapsto \ && \  IXI \ && \  \mapsto \ && \  XXX, &\\
&IZZ \ && \  \mapsto \ && \  IZI \ && \  \mapsto \ && \  ZZZ \ && \  \mapsto \ && \  ZII \ && \  \mapsto \ && \  ZZI \ && \  \mapsto \ && \  IIZ \ && \  \mapsto \ && \  ZIZ, &\\
&YII \ && \  \mapsto \ && \  YZX \ && \  \mapsto \ && \  IXY \ && \  \mapsto \ && \  ZIY \ && \  \mapsto \ && \  XYZ \ && \  \mapsto \ && \  IYI \ && \  \mapsto \ && \  YYY, &\\
&IYY \ && \  \mapsto \ && \  IZX \ && \  \mapsto \ && \  YYZ \ && \  \mapsto \ && \  ZXI \ && \  \mapsto \ && \  YYX \ && \  \mapsto \ && \  XIZ \ && \  \mapsto \ && \  YIY, &\\
&YIY \ && \  \mapsto \ && \  IYY \ && \  \mapsto \ && \  IZX \ && \  \mapsto \ && \  YYZ \ && \  \mapsto \ && \  ZXI \ && \  \mapsto \ && \  YYX \ && \  \mapsto \ && \  XIZ, &\\
&IYI \ && \  \mapsto \ && \  YYY \ && \  \mapsto \ && \  YII \ && \  \mapsto \ && \  YZX \ && \  \mapsto \ && \  IXY \ && \  \mapsto \ && \  ZIY \ && \  \mapsto \ && \  XYZ, &\\
&IXX \ && \  \mapsto \ && \  IIX \ && \  \mapsto \ && \  XXI \ && \  \mapsto \ && \  IXI \ && \  \mapsto \ && \  XXX \ && \  \mapsto \ && \  XII \ && \  \mapsto \ && \  XIX, &\\
&ZII \ && \  \mapsto \ && \  ZZI \ && \  \mapsto \ && \  IIZ \ && \  \mapsto \ && \  ZIZ \ && \  \mapsto \ && \  IZZ \ && \  \mapsto \ && \  IZI \ && \  \mapsto \ && \  ZZZ, &\\
&IIX \ && \  \mapsto \ && \  XXI \ && \  \mapsto \ && \  IXI \ && \  \mapsto \ && \  XXX \ && \  \mapsto \ && \  XII \ && \  \mapsto \ && \  XIX \ && \  \mapsto \ && \  IXX, &\\
&ZZI \ && \  \mapsto \ && \  IIZ \ && \  \mapsto \ && \  ZIZ \ && \  \mapsto \ && \  IZZ \ && \  \mapsto \ && \  IZI \ && \  \mapsto \ && \  ZZZ \ && \  \mapsto \ && \  ZII, &\\
&YYI \ && \  \mapsto \ && \  IXZ \ && \  \mapsto \ && \  YXY \ && \  \mapsto \ && \  XZZ \ && \  \mapsto \ && \  XZX \ && \  \mapsto \ && \  ZYY \ && \  \mapsto \ && \  ZIX, &\\
&XXI \ && \  \mapsto \ && \  IXI \ && \  \mapsto \ && \  XXX \ && \  \mapsto \ && \  XII \ && \  \mapsto \ && \  XIX \ && \  \mapsto \ && \  IXX \ && \  \mapsto \ && \  IIX, &\\
&IIZ \ && \  \mapsto \ && \  ZIZ \ && \  \mapsto \ && \  IZZ \ && \  \mapsto \ && \  IZI \ && \  \mapsto \ && \  ZZZ \ && \  \mapsto \ && \  ZII \ && \  \mapsto \ && \  ZZI, &\\
&IXI \ && \  \mapsto \ && \  XXX \ && \  \mapsto \ && \  XII \ && \  \mapsto \ && \  XIX \ && \  \mapsto \ && \  IXX \ && \  \mapsto \ && \  IIX \ && \  \mapsto \ && \  XXI, &\\
&ZIZ \ && \  \mapsto \ && \  IZZ \ && \  \mapsto \ && \  IZI \ && \  \mapsto \ && \  ZZZ \ && \  \mapsto \ && \  ZII \ && \  \mapsto \ && \  ZZI \ && \  \mapsto \ && \  IIZ, &\\
&IIY \ && \  \mapsto \ && \  YXZ \ && \  \mapsto \ && \  IYZ \ && \  \mapsto \ && \  XYX \ && \  \mapsto \ && \  YZZ \ && \  \mapsto \ && \  YIX \ && \  \mapsto \ && \  ZYX, &
\end{alignat*}
and for the $2_{4}14_{2} - 4_{3}6_{4}$ case we obtain
\begin{alignat*}{14}
&~~a \ && \  \mapsto \ && \  ~~b \ && \  \mapsto \ && \  ~~c \ && \  \mapsto \ && \  ~~d \ && \  \mapsto \ && \  ~~e \ && \  \mapsto \ && \  ~~f \ && \  \mapsto \ && \  ~~g, &\\
&IZI \ && \  \mapsto \ && \  ZZZ \ && \  \mapsto \ && \  ZII \ && \  \mapsto \ && \  ZZI \ && \  \mapsto \ && \  IIZ \ && \  \mapsto \ && \  ZIZ \ && \  \mapsto \ && \  IZZ, &\\
&IIY \ && \  \mapsto \ && \  YXZ \ && \  \mapsto \ && \  IYZ \ && \  \mapsto \ && \  XYX \ && \  \mapsto \ && \  YZZ \ && \  \mapsto \ && \  YIX \ && \  \mapsto \ && \  ZYX, &\\
&XII \ && \  \mapsto \ && \  XIX \ && \  \mapsto \ && \  IXX \ && \  \mapsto \ && \  IIX \ && \  \mapsto \ && \  XXI \ && \  \mapsto \ && \  IXI \ && \  \mapsto \ && \  XXX, \\
&IZZ \ && \  \mapsto \ && \  IZI \ && \  \mapsto \ && \  ZZZ \ && \  \mapsto \ && \  ZII \ && \  \mapsto \ && \  ZZI \ && \  \mapsto \ && \  IIZ \ && \  \mapsto \ && \  ZIZ, \\
&XYY \ && \  \mapsto \ && \  XZI \ && \  \mapsto \ && \  YZY \ && \  \mapsto \ && \  ZXX \ && \  \mapsto \ && \  ZZX \ && \  \mapsto \ && \  XXZ \ && \  \mapsto \ && \  ZXZ, &\\
&XXX \ && \  \mapsto \ && \  XII \ && \  \mapsto \ && \  XIX \ && \  \mapsto \ && \  IXX \ && \  \mapsto \ && \  IIX \ && \  \mapsto \ && \  XXI \ && \  \mapsto \ && \  IXI, &\\
&YXY \ && \  \mapsto \ && \  XZZ \ && \  \mapsto \ && \  XZX \ && \  \mapsto \ && \  ZYY \ && \  \mapsto \ && \  ZIX \ && \  \mapsto \ && \  YYI \ && \  \mapsto \ && \  IXZ, &\\
&ZIZ \ && \  \mapsto \ && \  IZZ \ && \  \mapsto \ && \  IZI \ && \  \mapsto \ && \  ZZZ \ && \  \mapsto \ && \  ZII \ && \  \mapsto \ && \  ZZI \ && \  \mapsto \ && \  IIZ, &\\
&IXI \ && \  \mapsto \ && \  XXX \ && \  \mapsto \ && \  XII \ && \  \mapsto \ && \  XIX \ && \  \mapsto \ && \  IXX \ && \  \mapsto \ && \  IIX \ && \  \mapsto \ && \  XXI, &\\
&IIZ \ && \  \mapsto \ && \  ZIZ \ && \  \mapsto \ && \  IZZ \ && \  \mapsto \ && \  IZI \ && \  \mapsto \ && \  ZZZ \ && \  \mapsto \ && \  ZII \ && \  \mapsto \ && \  ZZI, &\\
&ZZI \ && \  \mapsto \ && \  IIZ \ && \  \mapsto \ && \  ZIZ \ && \  \mapsto \ && \  IZZ \ && \  \mapsto \ && \  IZI \ && \  \mapsto \ && \  ZZZ \ && \  \mapsto \ && \  ZII, &\\
& IIX \ && \  \mapsto \ && \  XXI \ && \  \mapsto \ && \  IXI \ && \  \mapsto \ && \  XXX \ && \  \mapsto \ && \  XII \ && \  \mapsto \ && \  XIX \ && \  \mapsto \ && \  IXX, &\\
&ZII \ && \  \mapsto \ && \  ZZI \ && \  \mapsto \ && \  IIZ \ && \  \mapsto \ && \  ZIZ \ && \  \mapsto \ && \  IZZ \ && \  \mapsto \ && \  IZI \ && \  \mapsto \ && \  ZZZ, &\\
&YII \ && \  \mapsto \ && \  YZX \ && \  \mapsto \ && \  IXY \ && \  \mapsto \ && \  ZIY \ && \  \mapsto \ && \  XYZ \ && \  \mapsto \ && \  IYI \ && \  \mapsto \ && \  YYY, &\\
&IYI \ && \  \mapsto \ && \  YYY \ && \  \mapsto \ && \  YII \ && \  \mapsto \ && \  YZX \ && \  \mapsto \ && \  IXY \ && \  \mapsto \ && \  ZIY \ && \  \mapsto \ && \  XYZ, &\\
&YYX \ && \  \mapsto \ && \  XIZ \ && \  \mapsto \ && \  YIY \ && \  \mapsto \ && \  IYY \ && \  \mapsto \ && \  IZX \ && \  \mapsto \ && \  YYZ \ && \  \mapsto \ && \  ZXI. &
\end{alignat*}
We have also verif\/ied that all these replicas are `magic' in the same way as the conf\/iguration we started with, and each of them can thus be used as a proof of the KS theorem.
Similarly, we can get six magic replicas for any other W-A conf\/iguration by embedding it into the hexagon and `acting' on it by the automorphism in question.

\section{Conclusion}\label{section4}

Employing the structure of the split Cayley hexagon of order two, the smallest non-trivial generalized hexagon and a distinguished subgeometry of the symplectic polar space $W(5,2)$ of the three-qubit Pauli group, we got an intriguing f\/inite-geometric insight into the nature of a couple of `magic' three-qubit conf\/igurations proposed recently by Waegell and Aravind~\cite{wa}, namely the $18_{2} - 12_{3}$ and $2_{4}14_{2} - 4_{3}6_{4}$ ones. Either of the two conf\/igurations was found to be
uniquely extendible into a geometric hyperplane of the hexagon, this being, respectively, of type  ${\cal V}_{22}(37; 0, 12, 15, 10)$ and ${\cal V}_{4}(49; 0, 0, 21, 28)$ in the classif\/ication of Frohardt and Johnson~\cite{fj}. Moreover, an automorphism of order seven of the hexagon gave birth, for either of the two, to six more replicas, each having the same `magical' nature as the parent one.

It is important to emphasize that this is, to our best knowledge, only the second instance where geometric hyperplanes are related to certain `magic' conf\/igurations of observables. The f\/irst instance
was the Mermin square of two-qubits, which was recognized to be a full geometric hyperplane of the generalized quadrangle of order two, GQ$(2,2)$ (see, e.g., \cite{pla,ps,spp}). However, in a more general setting of quantum information theory, geometric hyperplanes have already entered the game through the concept of a Veldkamp space \cite{spph,vl}. Last but not least, we have to mention that this smallest split Cayley hexagon has also been found to play a role in the closely-related context of the so-called black-hole-qubit correspondence~\cite{lsv}; here also one of its geometric hyperplanes, the so-called distance-2-ovoid (of type ${\cal V}_{2}(21;21,0,0,0)$), was mentioned in connection with a~certain class of quantum codes.

That there is more behind this smallest split Cayley hexagon than meets the eye is also indicated by the following observation. One of the most symmetric, and most pronounced as well, three-qubit `magic' conf\/igurations is the so-called Mermin pentagram (see, e.g.,~\cite{mer,sl,wa}). Using computer, we have found that the full symplectic geometry of the three-qubit Pauli group,  $W(5,2)$ contains altogether 12\,096 copies of such a pentagram, this number being~-- remarkably~-- equal to the order of the {\it automorphism} group of our hexagon.

In their paper \cite[p.~7]{wa}, Waegell and Aravind stress that ``we have not displayed all the diagrams we have found, but only a representative sample''.  It would certainly be of great interest for us to become familiar with those yet unpublished and analyze them in the above-described fashion in order to see whether there is some underlying link between those that can uniquely be extended into a geometric hyperplane of our hexagon. We would, eventually, be most curious to see if each type of a geometric hyperplane can serve as a unique extension of some `magic' conf\/iguration(s), or whether this is a privilege for only some of them.

\subsection*{Acknowledgement}
This work was partially supported by the VEGA grant agency project 2/0098/10.

\pdfbookmark[1]{References}{ref}
\LastPageEnding

\end{document}